# Deriving the Traveler Behavior Information from Social Media: A Case Study in Manhattan with Twitter

Zhenhua Zhang [1]
Institute for Transportation, Iowa State University, Ames, IA 50010
Email: zhenhuaz@iastate.edu

---

1 Corresponding author

**Abstract**

Social media platforms, such as Twitter, provide a totally new perspective in dealing with the traffic problems and is anticipated to complement the traditional methods. The geo-tagged tweets can provide the Twitter users' location information and is being applied in traveler behavior analysis.

This paper explores the full potentials of Twitter in deriving travel behavior information and conducts a case study in Manhattan Area. A systematic method is proposed to extract displacement information from Twitter locations. Our study shows that Twitter has a unique demographics which combine not only local residents but also the tourists or passengers. For individual user, Twitter can uncover his/her travel behavior features including the time-of-day and location distributions on both weekdays and weekends. For all Twitter users, the aggregated travel behavior results also show that the time-of-day travel patterns in Manhattan Island resemble that of the traffic flow; the identification of OD pattern is also promising by comparing with the results of travel survey.

# 1 Introduction

Twitter has gradually become as a viable and even primary data source for some transportation research and applications. Recent studies valued its easy accessibility, low cost, as well as the "Big Data" features and made several breakthroughs in traditional research fields including traffic accident detection(1)(2), traffic flow prediction(3)(4), travel behavior and pattern analysis(5), transportation planning(6), infrastructure management(7), crisis management(8), etc. Of all these studies, travel behavior analysis focuses on the spatial and temporal travel features according to the GPS locations of the Twitter users and gives important information to transportation planners to evaluate the traffic operation.

This paper explores how to use the geo-tagged tweets to extract useful time-of-day travel behaviors. The literatures of travel behavior analysis using social media tools thrive in recent years. Before that, studies mostly use smart card automated fare collection (AFC)(9), household travel survey(10), mobile phone data(11), etc. In comparison with these data sources, Twitter has its own advantages.

First, Twitter provides the high-resolution GPS locations of the users and those tweets are called "geo-tagged tweets". The GPS locations have been proved useful in many studies based on mobile phone data, portable GPS devices as well as other social media tools: Ghahramani et al.(12) found that GPS-based information appears to be suitable for unveiling the intercity travel behaviors. Other studies even showed that 2-3% penetration of cell phones in the driver population is enough to provide accurate measurements of the traffic flow(13) and these floating sensor measurements even provide additional assurance for data accuracy(7). Second, besides the high-resolution GPS data, there is nearly no survey and installation cost for large information retrieval from Twitter. Twitter information can be streamed in large quantities and is comparatively much larger than that from carry-on GPS devices. Also, the big data collection does not limit to certain routes or corridors and this makes Twitter a much more flexible data source than smart card. Another important feature is that Twitter reflects the travel preferences of certain groups of people within a geographic area. As people tweet spontaneously when they want to share something, there is no experimental or mission-oriented settings for the Twitter users. In comparison, the traveler information from household travel survey comes from the questionnaires(10). Important features of the data sources in previous travel behavior studies are shown in Table 1.

Table 1 Features of Twitter and other data sources in travel behavior studies

|  | Data size | Data size per capita | Cost | Traveler demographics | Spontaneity | Location resolution | Fixed routes |
|---|---|---|---|---|---|---|---|
| Twitter | Big | Big | Low | Less details | Yes | High | No |
| Mobile phone data | Big | Big | Low | Less details | Yes | Low | No |
| Household travel survey | Big | Small | High | Full details | No | High | No |
| Smart card AFC | Big | Small | Low | Less details | Yes | High | Yes |

| Carry-on GPS devices | Small | Big | High | Full details | No | High | No |
|---|---|---|---|---|---|---|---|

There are also some limitations for Twitter-based travel behavior analysis. For instance, Twitter is sometimes not posted at trip ends, thus one cannot put these large GPS stream data into direct use of the travel behavior analysis. Also, due to GPS errors, the same user can be observed in two distinct locations with significant distance between them in a very small time gap(12).

Fully considering the advantages and disadvantages, this paper explores using Twitter to extract travel behavior information, especially the origin-destination (OD) patterns and time-of-day variations of the zone-based travel. To demonstrate its usability, we did a case study in New York Metropolitan Area and mainly extracted the people's travel behavior in and out of the Manhattan Island. The rest of the paper is intentionally organized to deliver our research purposes: Section 2 describes the study area, Twitter data features and its demographics; Section 3 details the systematic method we employ in this study to extract the Twitter travel information; Section 4 fully discusses the individual and aggregated travel behavior features, especially Twitter's advantages and disadvantages, applicability and representativeness in studying the travel behavior. Section 5 concludes the paper by a series of thoughtful conclusions and discussions.

## 2 Data description

### 2.1 Twitter data features

The study area is selected in the Manhattan, NY and its six nearest counties (boroughs) including Richmond (Staten Island), Kings (Brooklyn), Queens, Bronx, Bergen and Hudson. Millions of people getting in and out of Manhattan create huge traffic burdens to ferries, tunnels and bridges that connect Manhattan to its surrounding areas. For instance, average monthly traffic tunnels and bridges alone amount up to 9.5 million and this remains unchanged for the past 7 years(14).

In this paper, we collect more than 12.4 million geo-tagged Twitter data by an open streaming API. There are more than 17,000 Twitter users involved and the observation period is from Oct. 12 2013 - Feb.12 2016. The name: "geo-tagged" indicates the tweets are coupled with GPS locations and these take no more than 5% of the non-geo-tagged tweets. Even though, the data size competes with that of the household travel survey(10) and smart card AFC(9). The bounding box is set to fully cover the 7 counties except the less populated parts of North Bergen. Due to the different population density, one can visualize tweeting frequencies in these counties in Figure 1.

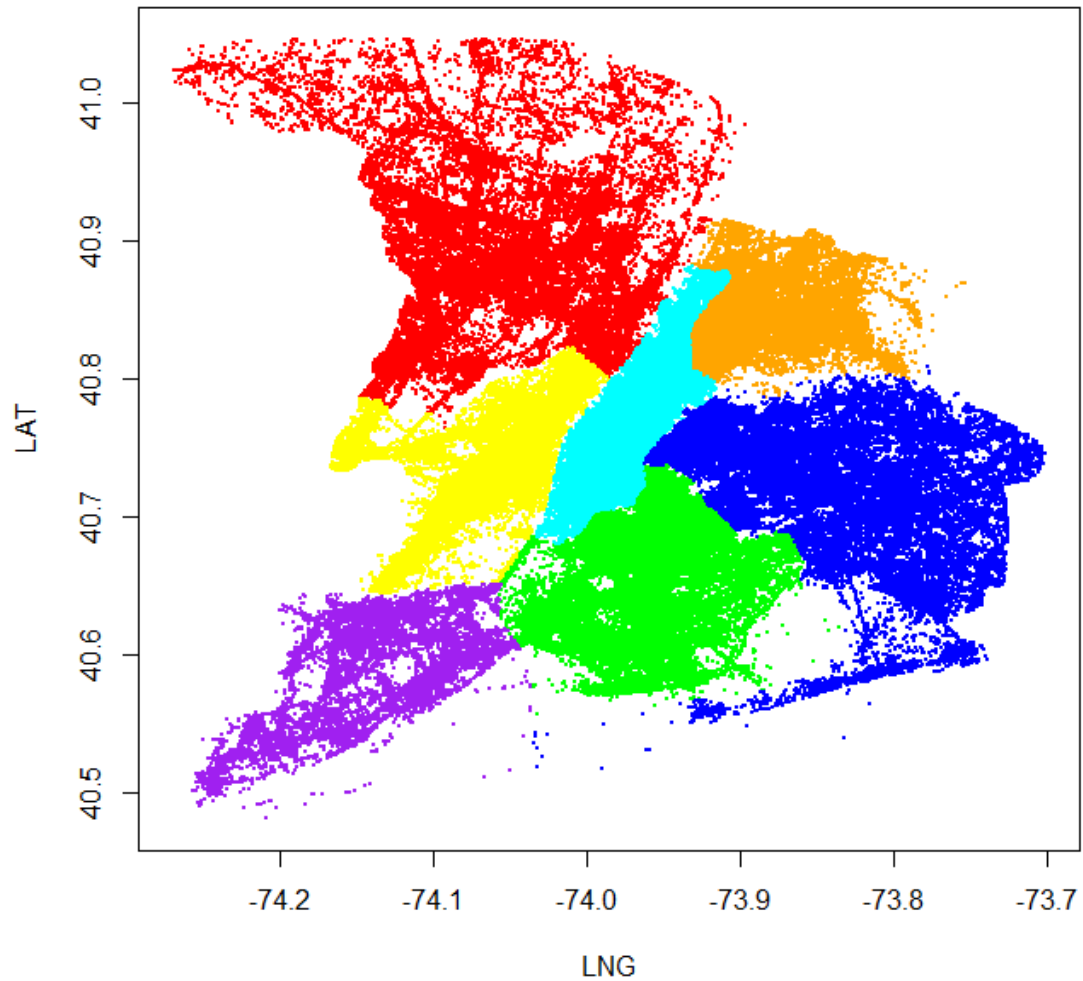

Figure 1 tweet locations in Manhattan, Richmond, Kings, Queens, Bronx, Bergen and Hudson from Oct. 12 2013 - Feb.12 2016. "LAT" = latitude and "LND" = "longitude".

We mainly use 5 attributes of the tweets: user ID, latitude, longitude, tweeting time, tweet posts. Their data formats and resolution is shown in Table 2.

Table 2 Summary of Twitter data features

|  | Example | Resolution | Range |
| --- | --- | --- | --- |
| User ID | 138987307 | / | 17,090 Twitter users |
| Longitude | -73.87725 | $10^{-6}$ | -74.3~-73.7 |
| Latitude | 40.99412 | $10^{-6}$ | 40.4~41.1 |
| Time | 8/2/2014 21:58 | second | Feb. 12 2015 - Feb.12 2016 |
| Tweet post | just posted a photo | / | / |

## 2.2 Twitter demographics

In traveler behavior analysis, one crucial information to the researchers is demographics of the travelers including their age, gender, education, etc. These information shows which groups of people conduct the trip, as well as the representativeness of the travel behavior results. Usually, different studies have different target groups. For instance, more than 60 percent of the interviewees in the 2009 household travel survey are more than 45 years old(10). Although we cannot directly interview the Twitter users in our collected datasets, there are still some open survey or investigation estimating their demographics. Table 3 listed education, age, gender and income distributions of Twitter users from comScore(15), Pew Research Center(16) and statista.com. Although most of the studies do not target the area of New York City, they still give good references to the related studies.

Table 3 Demographics of Twitter users

| Attribute | Source | | | | |
|---|---|---|---|---|---|
| Education | Pew Research Center | Less than high school | High school | Trade or some college | Bachelor's degree | Graduate school |
| | | 6% | 16% | 39% | 21% | 18% |
| Age | Statista.com | 18-24 | 25-34 | 35-44 | 45-54 | 55+ |
| | | 18.20% | 22.20% | 20% | 16.70% | 22.90% |
| Gender | Statista.com | Male | Female | | |
| | | 48% | 52% | | |
| Income | United States Census Bureau & Pew Research Center | 0-30K | 30K-50K | 50K-75K | 75K- |
| | | 23.15% | 8.35% | 13.51% | 54.98% |

From Table 3, one can see that average age of Twitter users are well below that of the travel survey. Twitter also has a good coverage of different groups of education, income level and gender.

## 3 Methods

This paper proposes a systematic method to extract the travel behavior results by pairing the sequential GPS location data into displacements of the tweet users. The process can be generalized by three major steps:

### 3.1 Featured Twitter user extraction

The data cleaning is to both extract useful location information and filter out GPS errors in the Twitter data, which can also be seen in the most of the previous GPS-based studies.

- First, in our datasets, the average number of tweets per user is about 730 but there exist large variations among Twitter users. We need to extract Twitter users who tweet often, higher than a threshold of 100. Our examination shows that there are more than 6,000 Twitter users who tweet more than the threshold during our observation period. Thus, the selected Twitter users in this study may be either the local residents near the Manhattan Area or frequent business travelers.
- Second, for each Twitter user, if we place their tweets by time order and calculate the time intervals and location differences between the consecutive tweets. We sometimes find a large location difference over a very small time interval and the average speed be higher than 100 mph. These may be due either to the GPS errors or special travel modes (e.g. sightseeing helicopter). These data are deleted as they are not related to our travel behavior study.

## 3.2 Displacement extraction by time window

The second step is to set a time window for time difference between the starting and ending Twitter time. If the Twitter user tweets in one location and after 10 minutes tweets in another location, it is reasonable to say the user makes a displacement during that 10 minutes. Thus, we can extract the displacements from two consecutive time-ordered Twitter locations whose time difference lies within the time window. The extracted displacements have certain limitations: the travel time may be either underestimated or overestimated by Twitter displacements as illustrated in Figure 2; also, due to the GPS errors, the Twitter locations may not be the exact trip origins and destinations.

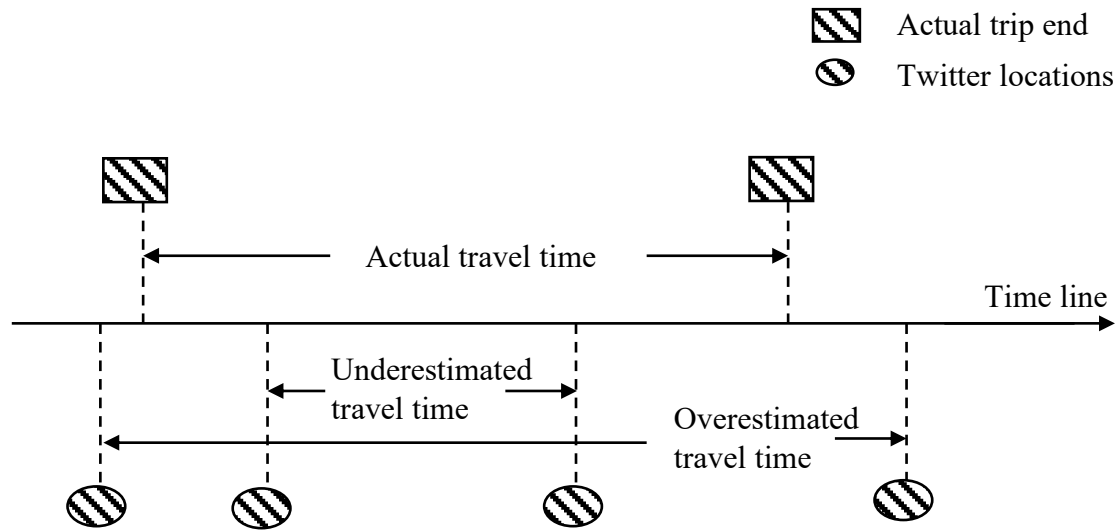

Figure 2 Errors of using displacement time interval in travel time estimation

## 3.3 Point-in-Polygon method

Without additional data source, it is not easy to overcome the limitations of Twitter time and locations as shown in Section 2.2. However, if we aggregate the Twitter locations into a large geographic scale (e.g. county) and extract the zone-based displacements, one may still extract accurate inter-county travel behavior.
- Under this condition, the GPS errors can be substantially reduced if we focus on the county-based travel.
- Given two time-ordered Twitter locations in different counties, one may estimate the border-crossing time. If the time window is set in 1 or 2 hours, the estimation can be accurate enough for time-of-day travel behavior analysis.

Thus, we need to create county labels for Twitter locations by the Point-in-Polygon method. For each county, we can use Google Map tools to manually select several locations as the boundary locations of the counties. In this case, we convert the location labelling problem into the Point-in-Polygon Problem by judging whether a given point (a Twitter location) lies inside or outside a polygon (county). To fast process the large Twitter datasets, we employ the methods by Bivand et al.(17). By setting the time window as 2 hours, we can finally obtain a large set of displacement results between the counties. The sample size is even larger than the household travel survey of the same area, which is published in 2009 as shown in Table 4. The great differences between the two data sources, as well as the features of their corresponding travel behavior results will be detailed in Section 4.

Table 4 Comparison of extracted displacements between Twitter and travel survey

|  | Twitter | Survey |
| --- | --- | --- |
| Number of travelers | 6,638 | 6,955 |
| Sample size | 96,471 | 13,674 |
| Average trips per traveler | 14.5 | 1.97 |
| Observation period | Oct. 12 2013 - Feb.12 2016 | 2010-2011 |

**4 Travel behavior from Twitter**

Based on the Twitter displacements, this section continues to analyze its travel behavior. In this section, we focus on two important features:
- Time-of-day and day-of-week travel behavior: this answers the time periods Twitter users are likely to make the movements, as well as weekday-weekend differences.
- OD travel behavior: this answers how the origin and destination Twitter locations distribute and the differences between inter-county (inter-city) travels in Manhattan area.

The numerical example starts by showing the individual pattern features we can obtain from Twitter; then we demonstrate and compare its travel behavior results in the region with those from two traditional data sources to prove its applicability.

**4.1 Individual travel behavior from two distinct groups**

Our initial examination can see two distinct groups of Twitter users based on the travel information. One may easily distinguish two groups by the tweeting times per user. If we order all the users from the highest tweeting times to the lowest, we can find that of all 6637 Twitter users, the top 1% users can be put into the first group who totally contribute nearly half of our displacements while the rest 99% in the second group do the other half. We can see that Twitter has a data collection mechanism combining both the experiment of distributed GPS device and household travel survey.

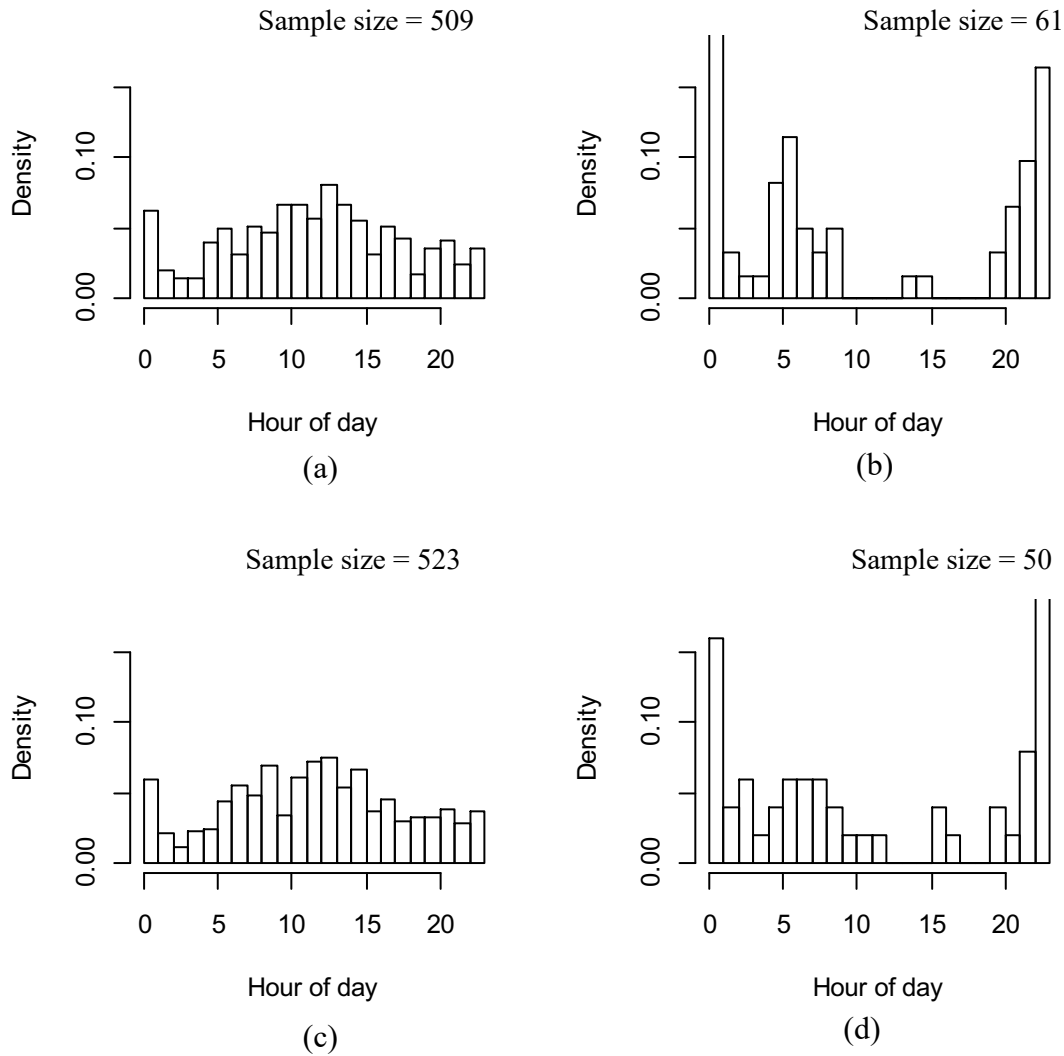

Figure 3 Hour-of-day travel behavior (ratio of travels) of one Twitter user from Manhattan to other counties (a) on weekdays; (b) on weekends; and from other counties to Manhattan (c) on weekdays; (d) on weekends

**The first group of Twitter users with high tweeting frequency**
In the first group, every Twitter user has a relatively large sample size of displacements. The individual travel behavior results can be unveiled based on the large datasets.

Figure 3 selects one Twitter user as an example and plots travel distribution over different hour of day on both weekdays and weekends. From the figures, one can easily figure out the peak hours of the user as well as the differences of peak hours between weekdays and weekends. One distinct pattern differences between weekdays and weekends can also be visualized.

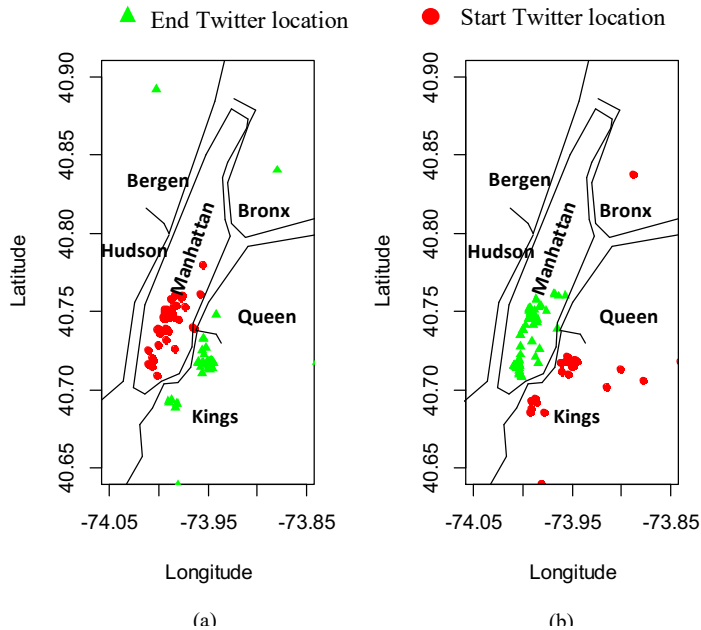

Figure 4 Historical travel behavior results for one Twitter user (a) from Manhattan to other counties; (b) from other counties to Manhattan

We can also see a clear patterns for Twitter locations of this user as shown in Figure 4, except for some occasional cases, one can see that the start and end locations can mostly be seen in a few geographic areas as shown in Figure 4. The Twitter user makes most of the movements between Manhattan and Kings: their destinations and origins center on a few fixed geographic areas considering the GPS errors. One can further check the year-to-year travel behavior changes over our observation period. If there is little fluctuations in the travel behavior over a long period, one can even make several meaningful presumptions. Overall, for the first group of Twitter users, based on their travel regularities both in time and locations. The continuous monitoring of the Twitter displacements resemble studying the travel behavior of the experiment participants who carry distributed GPS devices. They can both produce similar results exemplified in Figure 3 and 4. Both data sources help reveal the unique travel behavior features of individuals. However, compared with distributed GPS devices, travel behavior study based on Twitter excels in two ways:

- Twitter requires almost no installation or labor costs
- The operator can also obtain a much longer observation period.

**The second group of Twitter users with low tweeting frequency**

From the second group, we can only observe a limited set of displacements per user. In our displacement datasets, there are more than 5,000 Twitter users with less than 10 displacements in 3 years. The Twitter users with low tweeting frequency may be either a visitor or the local residents who do not tweet much.

Thus, the travel behavior extraction from these users resemble that from the household travel survey, in which the survey conductors aim to collect travel information from as many respondents as possible. In 2009 National Household Travel Survey, average size of travel records per respondent is no more than 10 except those with distributed GPS devices(10). Besides this similarity, Twitter overcomes some problems that exist in the household survey:

First, for the travel information part, the respondents in the survey are likely to actively report formal travel like commuting travel or work trips and less likely to give information like leisure walk or night travel. In comparison, the travel behavior from Twitter are collected and extracted passively. This is one advantage of Twitter travel behavior.
- Second, for the survey representativeness part, as the survey targets mostly on local residents, it may overlook the travel behavior of the passengers or tourists in the study area. Thus the results based on survey may lack representativeness as there are more frequent tourists or passengers in the Manhattan Area than other areas in each year. In comparison, similar problems in Twitter-based travel behavior will be relieved.

Above all, based on the tweeting frequency, locations, as well as the displacements, we can find the travel behavior features of the local commuters (in the first group), frequent or infrequent travelers. On one hand, this is good because Twitter covers a wide variety of people; on the other hand, even though we have the approximate distribution of Twitter users' age, gender, etc., we are still not sure whether they are local residents or tourists. Thus, there may also be bias problems in which Twitter may over-represent certain groups of people. The features of Twitter travel behavior will be further investigated and compared in the following aggregated results.

## 4.2 Aggregated travel behavior

This section continues with aggregated travel behavior from Twitter and still focuses on two important features: Time-of-day and day-of-week and OD travel behavior.

The first example is the travel behavior between Manhattan and Hudson. These two densely-populated counties are divided by Hudson River and are connected by several ferries, tunnels and bridges. This subsection mainly compares the time-of-day traffic patterns of Twitter, household travel survey and traffic flow between the two counties. For the traffic flow data, as it is difficult to count the exact traffic flow information of all travel modes, the traffic flow data in Lincoln Tunnel(14) is used instead. For travel survey data, the trips between Manhattan and Hudson can be extracted according to the origin and destination counties reported by the survey respondents. It is also worth mentioning that we compare the proportion of the traffic of three data sources over different hour periods instead of the traffic counts. The comparison in Figure 5 shows several interesting findings of the time-of-day travel behavior:

**4.2.1 Aggregated time-of-day travel behavior**

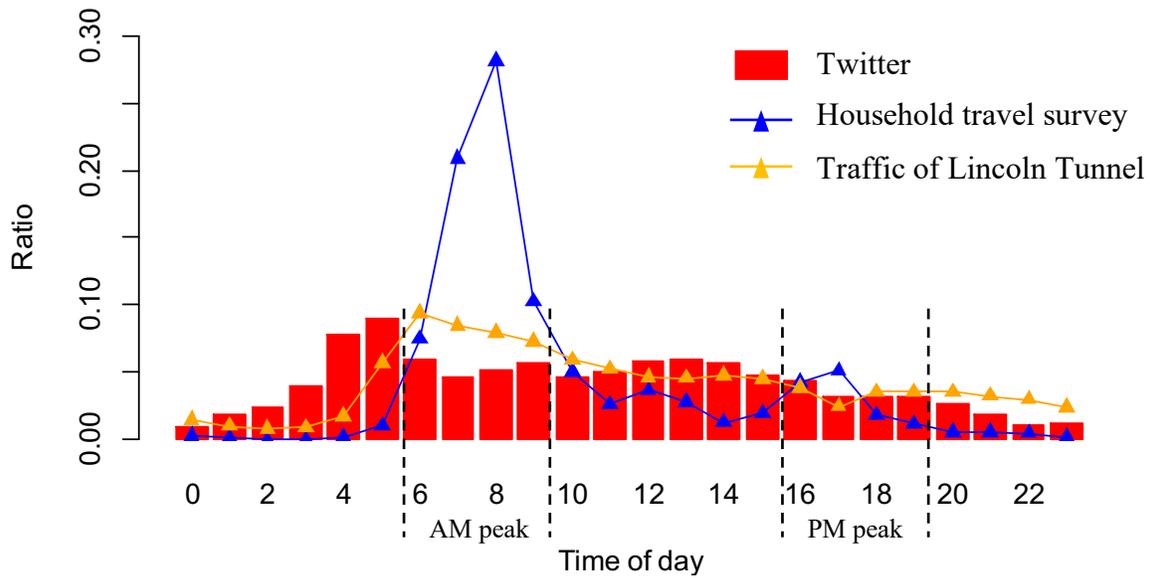

(a)

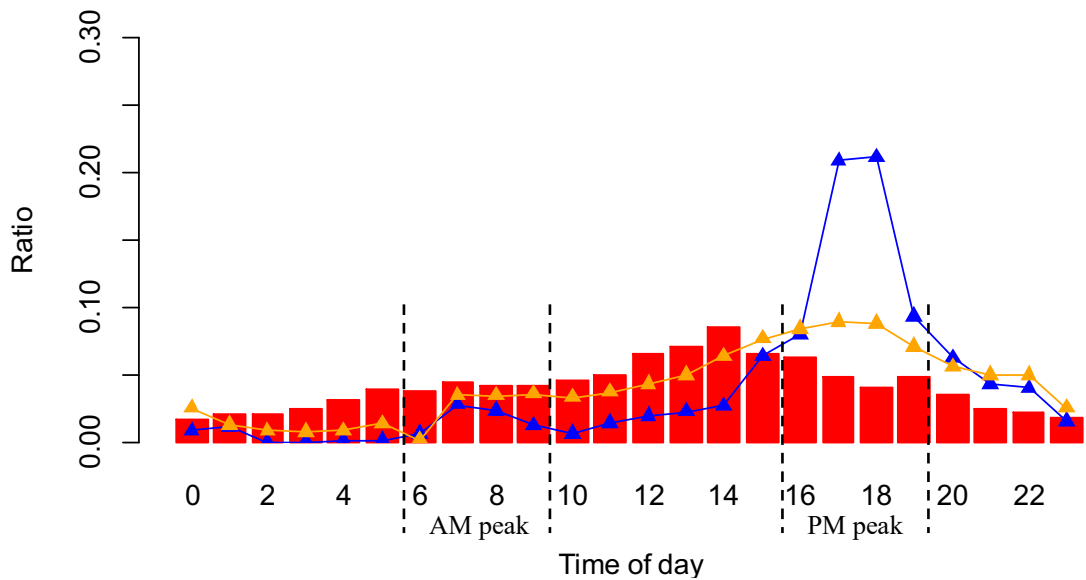

(b)

Figure 5 Aggregated time-of-day travel behavior results of Twitter, household travel survey and traffic flow ratio in Lincoln Tunnel (a) from Hudson to Manhattan; (b) from Manhattan to Hudson

**Finding 1: AM and PM peak in Twitter traffic patterns**

The usual AM peak and PM peak should happen from 6:00 a.m. to 9:00 a.m. and from 4:00 p.m. to 7:00 p.m., respectively. The AM peak happens on traffic from Hudson to Manhattan while PM peak does from Manhattan to Hudson. However, Twitter travel behavior of both directions do not show the expected AM and PM peak. The same problem happens in studies which employs Twitter to identify the longitudinal travel behavior (5).

Possible explanation is that Twitter travel behavior are extracted from tweets and people are more likely to tweet during leisure time. Therefore, for local residents there may not be so many tweets during AM or PM periods when people are busy with driving or commuting to work; and for tourists or passengers, there is also no AM or PM peak because they usually have flexible schedules. Even though, Twitter still sees the increase of border crossing behaviors before 6:00 a.m. from Hudson to Manhattan in Figure 5 (a), and before 4:00 p.m. from Manhattan to Hudson in Figure 5 (b). This trends potentially can be used to infer the peak hour travel behavior in the future studies

**Finding 2: the representativeness of Twitter travel behavior**

Figure 5 also shows an exaggerated AM and PM peak effects for the household travel survey. This coincides our assumption when comparing the travel behavior results between the survey respondents and the second group of Twitter users: survey respondents are more likely to report formal commuting or work travel and makes the travel behavior results biased. In contrast, Twitter travel behavior are more alike the border crossing traffic patterns in Lincoln Tunnel. Also, besides tunnels, there are also bridges and ferries that connect Manhattan to other counties. Therefore, the total traffic is usually difficult to collect. In this way, Twitter traffic patterns may be a viable supplement after some calibrations.

**4.2.2 Aggregated origin-destination patterns**

The second example focuses on the OD patterns which are important for traffic planning and travel demand forecasting. The OD patterns in this section mainly refer to those travels coming in and out of Manhattan and they are built on the 3-year Twitter displacement information. The OD estimation based on Twitter are alike those using mobile phone data or GPS-based devices in some literatures where the results have been validated by the temporal and spatial distributions of trips reported in local and national surveys (19). Similarly, in Figure 6, The Twitter travel behavior results are compared with the household travel survey in 2009. Note that aggregated OD patterns in Figure 6 compares mainly traffic proportions between counties instead of traffic because the traffic data in the bridges and tunnels connecting Manhattan Island are not easy to obtain.

However, from Figure 6, one can still see that the proportion of inter-county travel in Manhattan Area from Twitter are similar to that in the household travel survey and this may indicate that Twitter captures similar travel mechanism as that of the survey. Given that survey takes much more time and money, Twitter will be a viable tool to monitor the longitudinal OD demand changes.

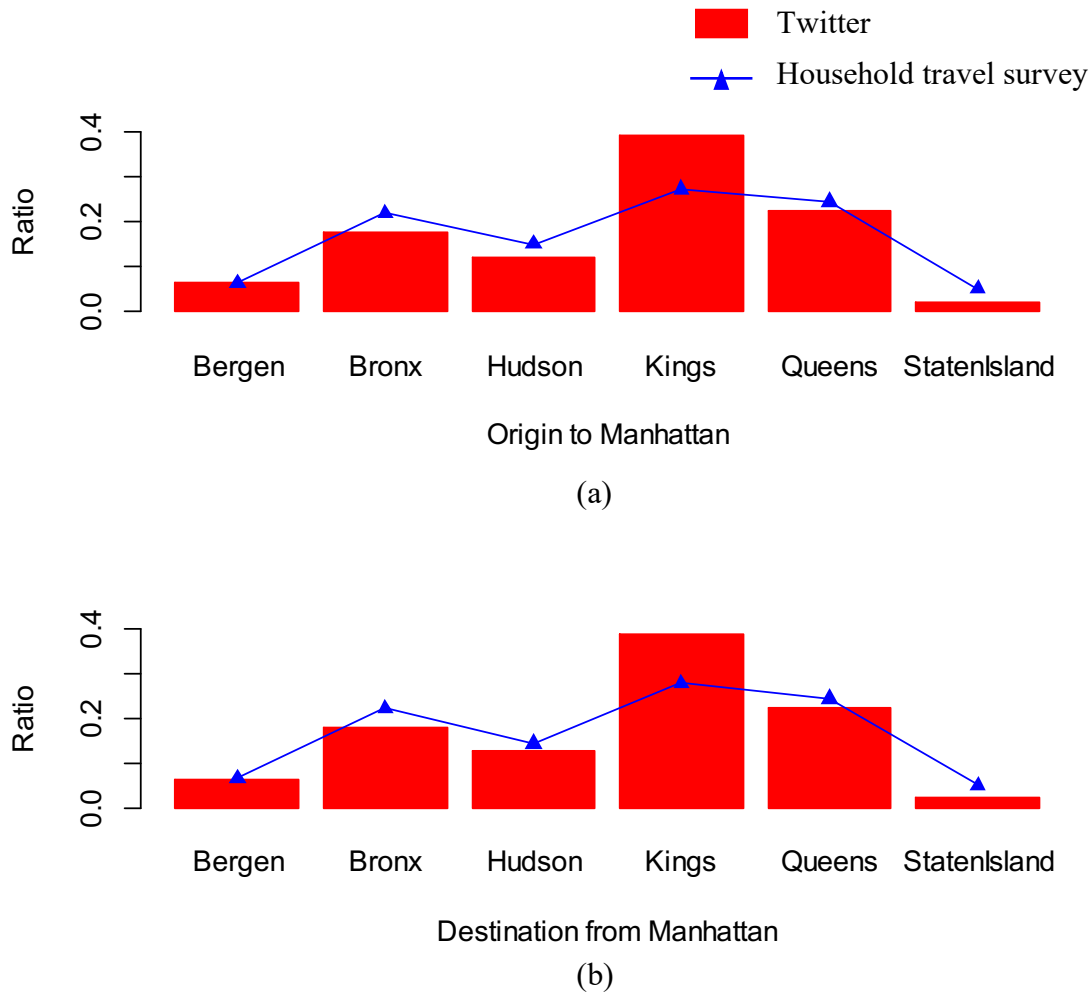

Figure 6 Aggregated origin-destination travel behavior of Twitter, household travel survey (a) from other counties to Manhattan; (b) from Manhattan to other counties.

## 5 Conclusions and discussions

This paper proposes a systematic method to extract useful travel behavior information from Twitter. The travel behavior results are fully detailed and discussed including the demographics of Twitter users, the representativeness, advantages and disadvantages compared with household travel survey and traffic data. Two important findings of Twitter-based travel behavior can be generalized as follows:

    First, based on the survey by comScore, Pew Research Center and statista.com, Twitter has a quite unique demographics as compared with travel survey. The extracted Twitter displacements can also be divided into two distinct groups and the data sizes of both are almost the same:

- Twitter users in the first group have a large tweet size per person and each of them shows unique time-of-day and location feature; and these travel behavior resemble

- those of the local residents. The data collection mechanism in this group is similar to that of household travel survey.
- Each Twitter user in the second group has limited Twitter displacements over 3-year observation period; and their location and time-of-day features are random and their travel behavior. A number of them may be those of the tourists of passengers and the data collection is similar to the experiments using carry-on GPS devices.

Second, from Twitter, we can also find the aggregated time-of-day and OD patterns. In the case studies in Manhattan, the time-of-day travel behavior from Twitter shows a clear increase of border crossing before AM and PM peaks but no apparent peak hour effects; By comparing with the traffic data in Lincoln Tunnel and the travel survey in Manhattan and Hudson, we can see that Twitter travel behavior complement the bias of travel survey and can be put into use after the peak hour effects are calibrated against the real traffic volume. Besides, the OD pattern identification derived from Twitter matches the pattern obtained from the travel survey.

Employing Twitter to monitor both the individual and aggregated travel behavior are similar to that of GPS-based devices or mobile phones, except that Twitter user has its own user groups which distinguish it from other data sources. By demonstrating the potentials of Twitter, one may also find some drawbacks: Twitter may also suffer from sampling bias problems; also, even though Twitter gives a good estimation of traffic distributions of origin and destinations, it cannot estimate the traffic volume. However, This bias is becoming less severe as social media users are growing making the sample a close representative of the population (20). The Twitter-based travel behavior results bring insight to the travel behavior analysis and will gradually complement the current travel studies.

**ACKNOWLEDGEMENT**

This study was partially supported by National Science Foundation award CMMI-1637604.

**Literatures**

1. Gu Y, Qian ZS, Chen F. From Twitter to detector: Real-time traffic incident detection using social media data. Transp Res Part C Emerg Technol. 2016;67:321–42.
2. Zhang Z, He Q, Gao J, Ni M. A deep learning approach for detecting traffic accidents from social media data. Transp Res Part C Emerg Technol. 2018;86.
3. Ni M, He Q, Gao J. Using social media to predict traffic flow under special event conditions. In: The 93rd Annual Meeting of Transportation Research Board. 2014.
4. Jiang X, Zhang L, Chen XM. Short-term forecasting of high-speed rail demand: A hybrid approach combining ensemble empirical mode decomposition and gray support vector machine with real-world applications in China. Transp Res Part C Emerg Technol. 2014;44:110–27.
5. Zhang Z, He Q, Zhu S. Potentials of using social media to infer the longitudinal travel behavior: A sequential model-based clustering method. Transp Res Part C Emerg Technol. 2017;85.
6. Camay S, Brown L, Makoid M. Role of social media in environmental review process of national environmental policy act. Transp Res Rec J Transp Res Board. 2012;(2307):99–107.
7. Bar-Gera H. Evaluation of a cellular phone-based system for measurements of traffic speeds and travel times: A case study from Israel. Transp Res Part C Emerg Technol. 2007;15(6):380–91.
8. Zhao X, Zhan M, Jie C. Examining multiplicity and dynamics of publics' crisis narratives with large-scale Twitter data. Public Relat Rev. 2018;44(4):619–32.
9. Kieu L-M, Bhaskar A, Chung E. A modified density-based scanning algorithm with noise for spatial travel pattern analysis from smart card AFC data. Transp Res Part C Emerg Technol. 2015;58:193–207.
10. Santos A, McGuckin N, Nakamoto HY, Gray D, Liss S. Summary of travel trends: 2009 national household travel survey. 2011.
11. Sadeghvaziri E, Rojas IV MB, Jin X. Exploring the potential of mobile phone data in travel pattern analysis. Transp Res Rec. 2016;2594(1):27–34.
12. Ghahramani N, Brakewood C, Peters J. An Exploratory Analysis of Intercity Travel Patterns Using Backend Data from a Transit Smartphone Application. 2017.
13. Herrera JC, Work DB, Herring R, Ban XJ, Jacobson Q, Bayen AM. Evaluation of traffic data obtained via GPS-enabled mobile phones: The Mobile Century field experiment. Transp Res Part C Emerg Technol. 2010;18(4):568–83.
14. Warf B. The Port Authority of New York-New Jersey. Prof Geogr. 1988;40(3):288–97.
15. Adam L, Andrew L. 2016 U.S. Cross-Platform Future in Focus [Internet]. 2016. Available from: http://www.comscore.com/
16. Duggan M, Brenner J. The demographics of social media users, 2012. Vol. 14. Pew Research Center's Internet & American Life Project Washington, DC; 2013.
17. Bivand RS, Pebesma EJ, Gómez-Rubio V, Pebesma EJ. Applied spatial data analysis with R. Vol. 747248717. Springer; 2008.